# Single-Crystal Growth of a FeCoCrMnAl High-Entropy Alloy

M. Feuerbacher[*], E. Würtz, C. Thomas

*Peter Grünberg Institut PGI-5, Forschungszentrum Jülich GmbH, D-52425 Jülich, Germany*

**Abstract**
We have grown a single crystal of the equiatomic FeCoCrMnAl high-entropy alloy. The crystal was grown by means of the Bridgman technique and is about 1 cm in diameter and 6.6 cm in length. X-ray Laue images taken at various positions on the surface are sharp and mutually consistent, reflecting its single-crystal nature. We thus report on the first successful growth of a single crystal of a high-entropy alloy with a volume of the order of cubic centimeters. The material has a microstructure consisting of B2 inclusions in a body-centered cubic matrix on the 10 nm length scale.

**Introduction**
High-entropy alloys (HEAs) constitute a novel field in materials science, increasingly attracting research interest. Yeh [1] was the first to forward the basic idea to study alloys composed of multiple principal elements, solidifying as metallic solid solutions on a simple crystal lattice. Usually, these alloys contain 4 to 9, occasionally up to 20 components [2]. As yet, several alloy systems in which HEAs exist have been reported, and HEAs with body-centered (bcc), face centered cubic (fcc), and hexagonal [3] structures have been found.
In this paper, we report on the single-crystal growth of a HEA. We approached the system FeCoCrMnAl, in which Pradeep et al. recently discovered a novel HEA phase [4]. In the literature, there exists only one previous article addressing single-crystal growth of a HEA [5]. The single crystal described in that paper, however, was fairly small, having a volume of about 0.35 cm$^3$. In the present paper, we explicitly address the growth of single crystals above a critical volume of 1 cm$^3$, which allows for the provision of single-crystalline samples for experimental techniques demanding large sample volumes, such as e.g. neutron diffraction.

**Methods**
An equiatomic melt of high-purity elements Fe, Co, Cr, Mn, and Al was produced using a levitation furnace. The alloy was remelted several times in order to ensure homogenization, and eventually cast into a conical mold. Single-crystal growth was carried out according to the Bridgman technique. The ingot was fitted in a cylindrical, slightly tapered alumina crucible of about 9 cm length and an internal diameter between 9 and 11 mm. The crucible was inserted in a molybdenum tube acting as a susceptor for high-frequency heating and placed on a moveable rod equipped with a cold finger to create a defined and steep temperature gradient. At the beginning of the growth process, the crucible was fully inserted into the susceptor. The temperature was measured pyrometrically through holes drilled into the susceptor tube. In the lower part of the crucible the temperature was set to 1530 °C and kept constant during the growth process. Since the liquidus temperature (determined using a Setaram Labysys DSC16) of the material is about 1400 °C, the entire ingot is initially molten and the overheating leads to further homogenization of the melt.

---

[*] Corresponding autor. Email m.feuerbacher@fz-juelich.de

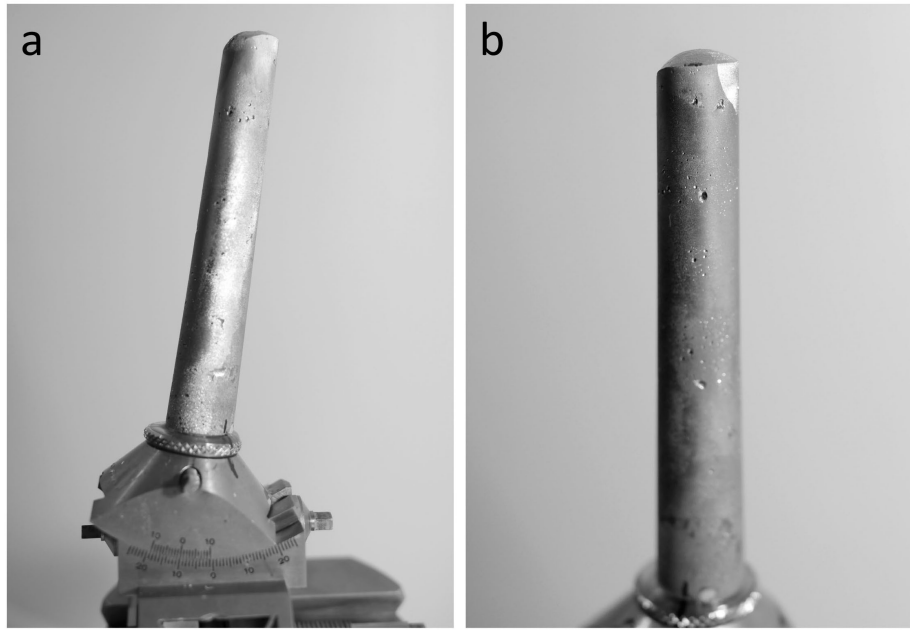

Fig. 1: Photograph of the Bridgman crystal grown mounted on a goniometer head (a). (b) The tip of the crystal oriented such that a small grain in the upper right corner is oriented in a specular direction, thus appearing bright.

Solidification was effected by lowering the crucible relative to the susceptor, i.e. by slowly moving the melt out of the hot zone, starting with the tip. We used a lowering velocity of 5 mm/h. The growth process was carried out under argon atmosphere of 650 mbar.

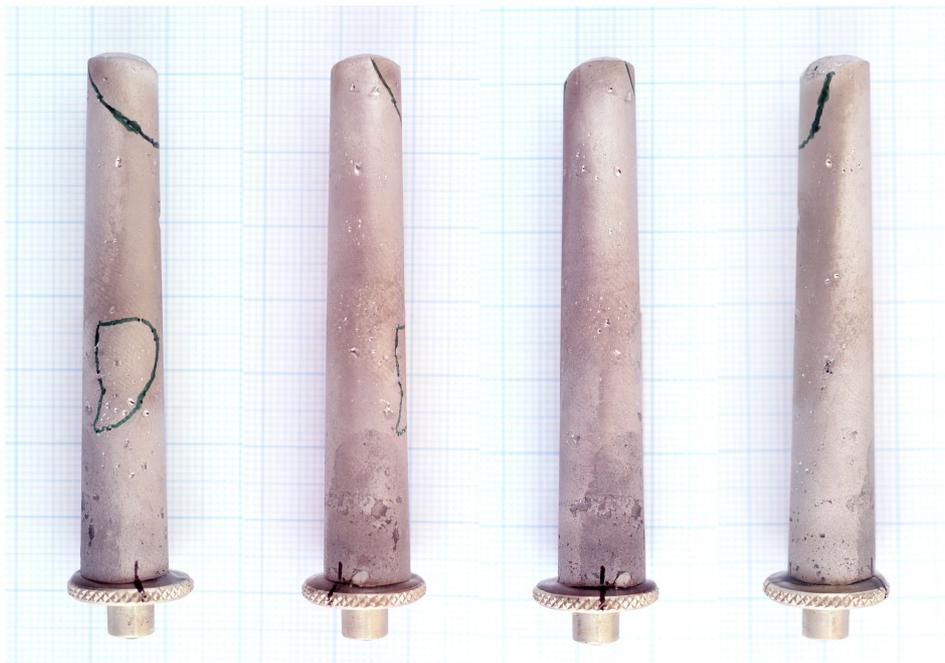

Fig. 2: Photographs of the grown crystal from several directions, so that its entire surface can be seen. The grain boundaries are traced on the surface using a marker pen.

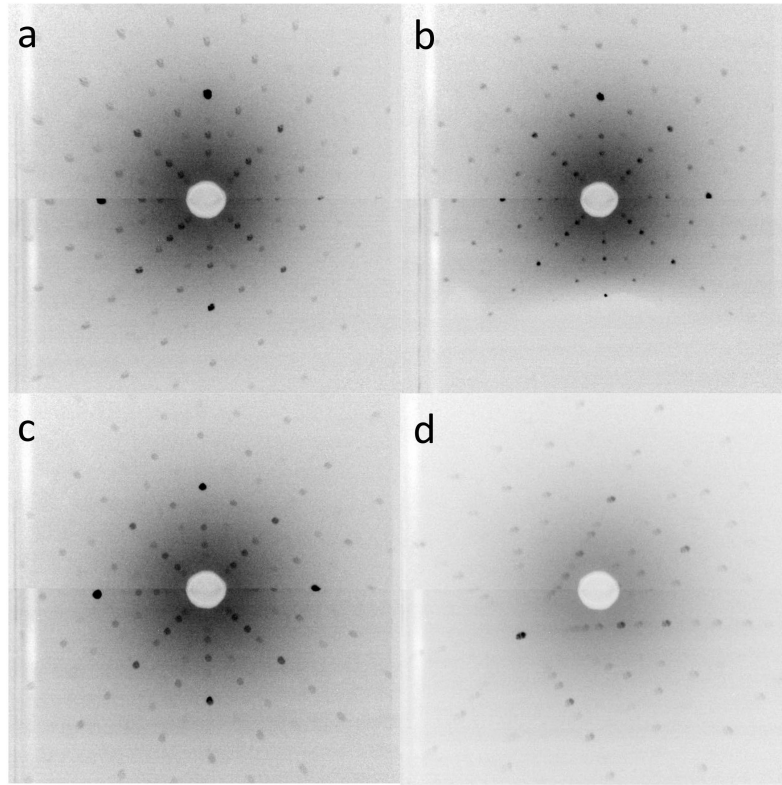

Fig. 3: Back-reflection Laue images of the (1 0 0) plane. (a) At the uppermost part of the primary grain. (b) At the lowermost part of the primary grain. (c) In the upper part on the opposite side of the crystal. (d) On the secondary grain in the centre of the crystal.

After a first visual inspection, the extension of the single grains and the primary orientation was determined using a Philips Mikro X-ray Laue apparatus in back-reflection geometry. The images were collected digitally using a Photonic Science X-ray Laue Imaging System. For further characterization, slices perpendicular to the [1 1 0] direction were extracted from the crystal by means of spark erosion. Scanning electron microscopy (SEM) was carried out using a JEOL 840 microscope, equipped with an EDAX energy-dispersive X-ray (EDX) system. Sample preparation for transmission electron microscopy (TEM) was carried out by subsequent grinding, dimpling and argon-ion milling. TEM was performed using a Philips CM20 transmission electron microscope operated at 200 kV.

**Results and discussion**
Fig. 1a shows the Bridgman crystal grown mounted on a goniometer head. Its outer appearance corresponds to the inner shape of the alumina crucible. The length of the crystal is 6.6 cm, the diameter is 11 mm at the bottom and 9 mm at the top, i.e. its total volume amounts to 5.1 cm$^3$. On the surface of the crystals a number of small cavities are visible.
The surface of the crystal is covered with microfacets, causing specular light reflection, thus leading to brighter or darker appearance of the crystal surface depending on its orientation. Fig. 1b shows the tip of the crystal at an orientation, where a small grain in the upper right corner is oriented in a specular direction, thus appearing bright in the photograph, while the rest of the crystal appears darker. Optical inspection rotating the crystal under incident light, accordingly allows for a first judgment of the extension of the grains and/or the presence of secondary grains. Fig. 2 shows the grown crystal from several directions, so that its entire surface can be seen. The grain boundaries located according to visual inspection are drawn on the surface using a marker pen. The whole crystal consists of three grains. In the upper part, a small wedge shaped grain is present, and in its center a second, irregularly shaped grain is present. For the entire remaining bulk, no changes in the appearance of the specular reflections is seen, it visually appears as a single grain. We will refer to this part as the primary grain of the crystal.

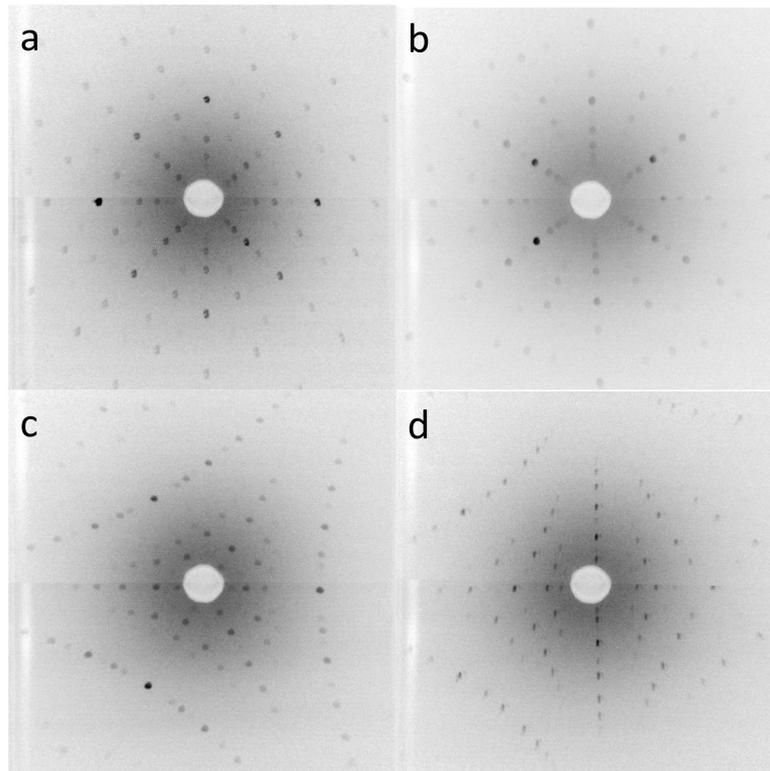

Fig. 4: Back-reflection Laue images taken in the center of the primary grain. (a) (1 0 0) plane. (b) (0 1 -1) plane. (c) (1 1 -1) plane. (d) (2 1 -1) plane.

The crystal was carefully inspected by X-ray Laue diffraction at various positions on its surface. Fig. 3a to c show back-reflection Laue images of the (1 0 0) plane in the primary grain of the crystal. Fig. 3a was taken at the uppermost part. Clearly, the fourfold symmetry of the pattern is seen. Fig. 3b is a Laue image taken at the lowermost part of the crystal. No changes of the crystal orientation were made; the crystal was only translated linearly to expose a spot in the lower part to the beam. The Laue image obtained at this position exactly corresponds to that taken in the upper part of the crystal. Note that in the lower quarter of the Laue image shadowing occurs, which is due to the close proximity of the exposed crystal area to the mounting plate. The two positions at which Laue images were taken are 60 mm apart on the crystal surface. Diverse Laue images taken in between these two extreme positions were all identical. Fig. 3c was taken in the upper part of the crystal, but on the opposite side of the crystal and in the opposite direction with respect to the position of Fig. 3. Again, we find the same characteristic pattern of the (100) plane.

Fig. 3 d was taken in the region of the secondary grain in the center of the crystal. Here we obtain a different Laue image. The 1 0 0 spot is tilted to the lower left by about 17 degrees and rotated by about 53 degrees. This part of the crystal thus corresponds to a different grain, oriented arbitrarily with respect to the large primary grain, which corresponds to our conclusion from the visual inspection.

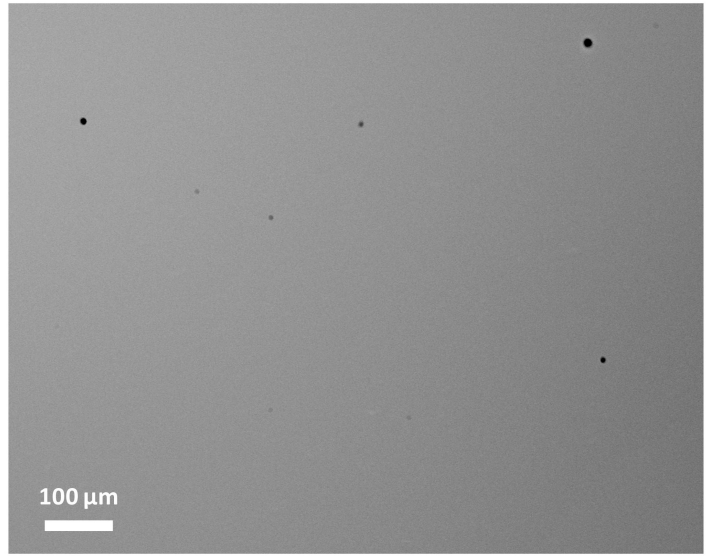

Fig. 5: Scanning electron micrograph of the material using a backscattered-electron detector.

Fig. 4 shows a series of Laue images taken around the central part of the crystal, all in the primary grain. The images were taken by rotating the crystal on a goniometer and capturing Laue images on the resulting different positions on its surface. All Laue images show sharp singular spots, and are consistent concerning their relative tilting angles. Fig. 4a is the fourfold pattern of the (1 0 0) plane. Fig. 4b is the twofold pattern of the (0 1 -1) plane and Fig. 4c the threefold pattern of the (1 1 -1) plane, which are tilted by 90 and 55 degrees with respect to (1 0 0), and 35 degrees with respect to each other. Finally, Fig. 4d shows the pattern of the (2 1 -1) plane, which was obtained after tilting by 35 degrees with respect to (1 0 0). The long axis of the crystal, i.e. the growth direction, is tilted by about 14 degrees with respect to the [0 1 1] direction.

These results demonstrate that the primary grain consists of one large single crystal. We can estimate the volumes of the two additional grains: The wedge shaped grain at the top of the crystal has a volume of about 0.24 cm³. Assuming that the visible edges of the secondary grain in the center of the crystal are directly connected, its volume can be estimated as 0.07 cm³. Hence, the volume of the primary grain is as large as 4.85 cm³.

Fig. 5 displays a scanning electron micrograph of the material using a composition-sensitive backscattered-electron detector. The image reveals that on this length scale, the sample is perfectly homogeneous. There are no composition variations and no dendrites, and no precipitates or secondary phases can be seen. A few dark spots on the image are visible, which are due to small pores that are present in the crystal at a very low density. EDX analysis averaged over the surface area reveals that the overall composition of the crystal grown is $Fe_{21.1}Co_{20.3}Cr_{21.2}Mn_{15.4}Al_{22.0}$, which is close to the nominal composition.

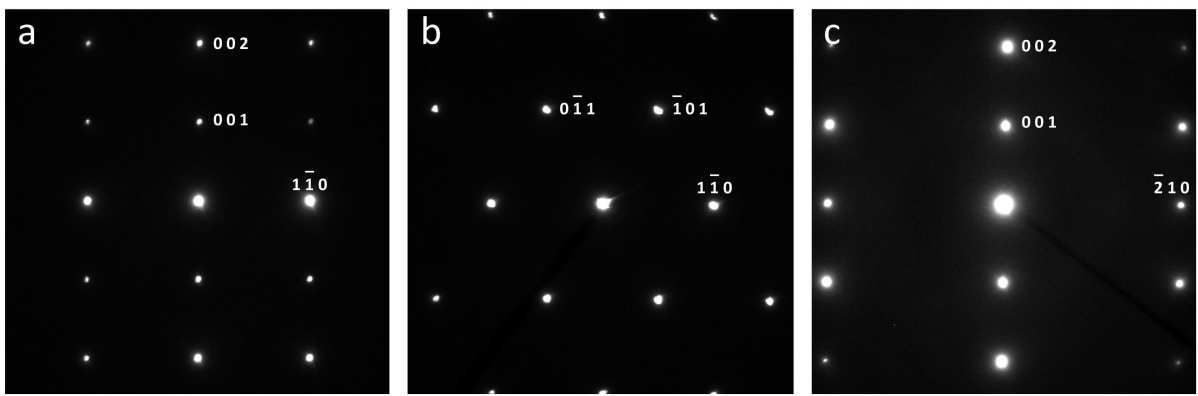

Fig. 6: Selected area diffraction patterns taken in the TEM. (a) [1 0 0] zone axis. (b) [1 1 1] zone axis. (c) [1 2 0] zone axis. Some reflections are indexed.

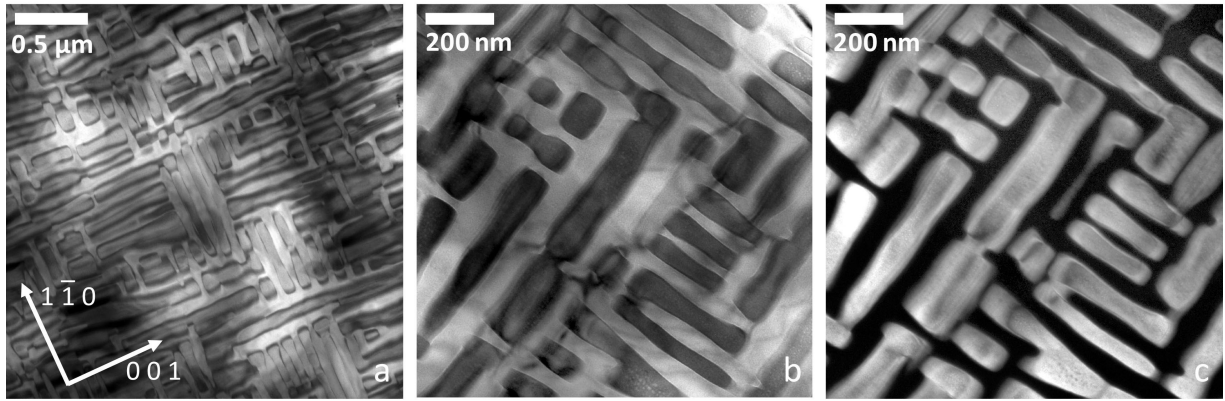

Fig. 7: TEM micrographs taken close to the [1 1 0] zone axis displaying the typical microstructure. (a, b) Bright-field images taken with the reflections (0 0 1) and (0 0 2) simultaneously excited. (c) Dark-field image of the region shown in (b) using the (0 0 1) reflection.

Fig. 6 shows selected-area electron diffraction patterns taken in the transmission electron microscope. The area contributing to diffraction is about 1 μm². Diffraction patterns corresponding to the [1 1 0], [1 1 1], and [1 2 0] zone axes are shown in panels a, b, and c, respectively. In the [1 1 0] pattern, the presence of (0 0 1) reflections is clearly visible. This reflection is forbidden in the bcc structure and reveals that the material has a B2 structure. The (0 0 1) reflections are also visible at the [1 2 0] zone axis.

Fig. 7a and b show bright-field TEM micrographs taken close to the [1 1 0] zone axis with the reflections (0 0 1) and (0 0 2) simultaneously excited. The images display the typical microstructure of the material on the submicron scale. The material consists of a matrix and inclusions, both existing at an approximately equal volume fraction. The inclusions are elongated and extend along the [0 0 1] and [1 -1 0] directions. While the length of the inclusions varies strongly, their width is rather uniform. The average width of the inclusions extending along the [1 -1 0] direction is 63.5 nm, and that of the inclusions extending along [0 0 1] is 90.8 nm.

Fig. 7c is a dark-field image using the (0 0 1) reflection of the exact area shown in Fig. 7b. Under these conditions the inclusions are bright while the matrix is dark. The (0 0 1) reflection is present in the B2 structure but forbidden in the bcc structure. Hence we can conclude that the inclusions possess B2 structure, while the matrix is bcc. The two-phase structure is thus similar to that found in equiatomic AlCoCrFeNi HEAs, where the presence of bcc and B2 phases was also found [6]. However, the matrix/inclusion setting is inverted: while the FeCoCrMnNi HEA presented in this study has a bcc matrix and B2 inclusions, in AlCoCrFeNi HEAs the matrix is B2 ordered and the inclusions are bcc. Further investigations employing analytical TEM are currently underway and will provide insight into the chemical nature of the inclusions and matrix.

Furthermore, during handling we noticed that the material is hard, brittle and ferromagnetic.

## Conclusions and outlook

In this paper we demonstrate the successful growth of a single crystal of a high entropy alloy of the equiatomic phase FeCoCrMnNi. The crystal grown consists of a primary grain and two small secondary grains. The volume of the primary grain is 4.85 cm$^3$. To the best of our knowledge this is the first report of the successful growth of a cubic centimeter sized single crystal of a high-entropy alloy.

The availability of single-crystalline sample materials is highly important for the determination of intrinsic materials properties without stray influences of the results by grain boundaries or other impurities. Furthermore, large samples are demanded by several characterization techniques such as neutron scattering, requiring sample volumes of the order of 1 cm$^3$, surface investigation techniques such as scanning tunneling microscopy combined with low-energy electron diffraction, and corrosion and tribology studies, requiring sample surface areas of several tens of mm$^2$ and lateral sample surface extensions of up to 10 mm. Further examples are low-frequency elastic constant measurements and anelasticity measurements with a torsion pendulum, etc. Neutron scattering for the determination of short- and medium range order is of particular importance in HEAs: most HEA phases include atom species which are closely neighbored in the periodic table, thus having very similar atomic numbers. As a result, X-ray and electron diffraction techniques will suffer from a lack of contrast and thus may be only of limited use for structural studies of these materials.

There is one previous publication claiming the single crystal growth of a high-entropy alloy by Ma et al. [5]. These authors report on the growth of a single crystal in the system CoCrFeNiAl$_{0.3}$ by means of a Bridgman technique. The crystal grown, however, is only 3 mm in diameter at a length of 50 mm. Its total volume is thus only 0.35 cm$^3$ and therewith only allows for a limited range of experimental approaches.

Our study shows that single-crystal growth of HEAs up to a size of several cm$^3$ is feasible. The size of the single crystal presented in this study is limited by that of the crucible used. The growth of even larger crystals, if required, seems achievable. In a next step, we will approach the growth of oriented single crystals using seeded Bridgman or the Czochralski technique. Furthermore, we are on the verge of developing single-crystal growth routes of related HEAs, such as face centered cubic FeCoCrMnNi and FeCoCrMnPd, in order to allow comparative studies of physical properties, which will allow deeper insight into the structure-property relations of this class of materials.

## Acknowledgments

The authors thank K.G. Pradeep and A. Marshal for disclosing their discovery of the FeCoCrMnAl HEA to us prior to publication.